\documentclass[12pt,cite,epsf,epsfig]{article}
\usepackage{epsfig}
\title{$t \to b W$ in NonCommutative Standard Model}
\author{Namit Mahajan\thanks{E--mail : nmahajan@mri.ernet.in}\\
	{\em Harish-Chandra Research Institute,} \\
	 {\em Chhatanag Road, Jhunsi, Allahabad - 211019, India.}}

\def\be{\begin{equation}}
\def\ee{\end{equation}}
\def\bea{\begin{eqnarray}}
\def\eea{\end{eqnarray}}

\begin{document}
%\doublespacing
\maketitle
%\large

\begin{abstract}
We study the top quark decay to b quark and W boson in the
NonCommutative Standard Model (NCSM).  The lowest contribution to the
decay comes from the terms quadratic in the matrix describing the
noncommutative (NC) effects while the linear term is seen to
identically vanish because of symmetry. The NC effects are found to be
significant only for low values of the NC characteristic scale.\\
%{\bf Keywords}: \\
\end{abstract} 

\begin{section}{Introduction}
\indent The belief behind the existence of the top quark, even before it was discovered \cite{discovery},
had its reasons buried deep in the consistency of the Standard Model (SM), whether it was due to requirements
of anomaly cancellation or to explain the precision electroweak data. The experimental evidence for the top quark
strengthened the three family structure of the SM and opened a whole new world of top-physics. The top quark is the
heaviest of all the fundamental particles known having a mass $174.3\pm 5.1$ GeV and contributes significantly to 
radiative processes. Such a large mass leads to a very short life time for the top quark ($\tau_{top}\sim 4\times
10^{-25}~$s). This number is roughly one order of magnitude smaller than the typical QCD hadronization time scale
($30\times 10^{-25}~$s). The top quark therefore decays before it can hadronise unlike the other quarks. This feature
offers a possibility of looking at a quark and its properties almost free of the QCD confinement. This makes the top quark
a wonderful laboratory for QCD studies.
To complete the picture
consistent with the SM, various properties of the top quark need to be confirmed. For a review on top quark and related issues
see \cite{review} and further references therein.\\ \\
\indent Within the SM, the top quark decays almost completely into a b quark and a W boson with the branching ratio
$BR(t\to b W)\sim 0.998$. Like the other particles, the charged interactions of the top quark in the SM are of V-A nature and
the strengths for the top quark going into a W boson and a lighter quark are directly proportional to the CKM mixing angles.
The top quark, in contrast to any other particle say a b quark, decays into a lighter quark (mostly a b quark) and an on-shell
W. This makes the helicity of the emitted W boson an important tool for determining many properties and testing the universal
nature of the V-A interactions. In the limit of massless b quark, the W cannot be right handed (or atleast such a helicity state
will be highly suppressed). The W helicity can be measured from the angular corelations of its decay products and thus offers a 
window to look for new physics beyond the SM as well. Due to the smallness of other CKM elements, other top decays fall into the 
category of rare decays and the branching ratios are very small. Top decays beyond the SM have been extensively studied. For example,
the top decay to charged Higgs and the experimental signatures has been discussed in detail in \cite{higgs} while the neutral current
decay into charm and two vector particles has been considered in \cite{diaz}. In \cite{stop} the decay into stop and neutralino has been
studied while the authors of \cite{tc2} consider the technicolour models and discuss related issues.\\ \\
 \indent The top quark properties play an important role in the electroweak physics. The large mass of the top quark
makes its role in the electrweak precision data fits much more pronounced than any other quark. Also the spin configuration
of the top quark in any process is very sensitive to new physics beyond the SM, particularly any anomalous coupling
other than allowed by the SM. The anomalous top quark couplings have been a very important and interesting area of activity.
These couplings can be probed directly at the colliders and indirectly via the rare decays of mesons \cite{colliders, rare}. 
It is expected that about $10^8$ top quark pairs pers year at the LHC will make the detailed study of the top quark couplings
possible and with great accuracy. Since the dominant decay channel for the top is decay into a b quark and a W boson, it is
not wrong to expect that the Wtb coupling will be measured with high precision. This coupling is proportional to the CKM element
$V_{tb}$ and enters the expressions for the falvour changing neutral current B-meson decays \cite{buchalla}. Therefore,
any anomalous contribution to the coupling should show up in the B-decays and should be measurable even before LHC begins.
Using the CLEO and LEP data the top couplings have been constrained \cite{constraint} and future experiments are expected to
improve these constraints. \\ \\
\indent In the present note we study the dominant decay mode of the top quark, $t \to b W$ in the context of the 
NonCommutative Standard Model (NCSM). The simple picture of space-time that we have in our minds is based on
the notion of space-time being described by a suitable manifold with the points on that manifold being labelled by
a countable number of real coordinates. For most of the practical purposes the space-time acts as a static background
on top of which the processes occur. However, it is believed that this naive picture must undergo a drastic change
when one probes very small distances. At those energy scales the classical notions of space-time seize to be the 
correct description of the world and modifications have to be incorporated to yield correct results. There is no clear cut
and unique solution to this puzzle and the kind of modifications required but a possible way to approach this problem is
to formulate physical theories on noncommutative (NC) space-times. This idea dates back to the work of Snyder \cite{snyder} 
where it was shown that the usual continuum space-time is not the only solution to the assumption that the spectrum of the
coordinates describing the space-time is invariant under Lorentz transformations and there exists a Lorentz invariant
space-time with an inherent unit of length. In such cases, the notion of a point is not well defined and the usual
commutation relations between the coordinates and momenta get an extra piece proportional to momentum value.
Therefore, for small enough momenta and energies, they just approach the usual quantum mechanical relations. A very strong
motivation to formulate quantum theory on a noncommutative space-time was to render the quantum field theory calculations
finite and free of infinities. But the success of the renormalization theory abdoned this approach altogether. The idea has
been remotivated because of some string theory results \cite{stringy}. Apart from being boosted by the string theory arguments,
field theories formulated on noncommutative space-times have very interesting features of their own. For a review
of noncommutative field theories (NCFTs) see \cite{ncreview}.\\ \\
\indent In order to describe a noncommutative space-time, the usual commutation property between the coordinates is
abandoned and replaced with the following commutator for the hermitian operators ${\hat{x}}^{\mu}$ \cite{filk}
\be
[{\hat{x}}^{\mu},{\hat{x}}^{\nu}] = i\Theta^{\mu\nu}
\ee
where $\Theta^{\mu\nu}$ is a constant, real and antisymmetric matrix and describes the noncommutativity.
This constant matrix can also be thought of as some background field relative to which the various 
space-time directions are distinguished. A priori there is no reason to assume that $\Theta^{\mu\nu}$ is a 
constant matrix but we consider it to be so in our study. Also, most of the studies involving field theories
on such spaces have been restricted to constant $\Theta^{\mu\nu}$. 
The theories on such a space-time clearly violate Lorentz invariance explicitly. Following the Weyl-Moyal correspondence \cite{moyal},
the ordinary product of two functions is replaced by the Moyal product (sometimes also called star product and denoted by
$\ast$) which takes the form
\bea
f(x)\ast g(x) &=& \Bigg[exp\Bigg(\frac{i}{2}\Theta^{\mu\nu}{\partial_{\eta}}_{\mu}{\partial_{\zeta}}_{\nu}\Bigg)f(x+\eta)
g(x+\zeta)\Bigg]_{\eta=\zeta = 0} \\ \nonumber
&=& \sum_{n=0}^{\infty}\Bigg(\frac{i}{2}\Bigg)^n\frac{1}{n!}[\partial_{\mu_1}\partial_{\mu_2}....\partial_{\mu_n}f(x)]
\Theta^{\mu_1\nu_1}\Theta^{\mu_2\nu_2}....\Theta^{\mu_n\nu_n}[\partial_{\nu_1}\partial_{\nu_2}....\partial_{\nu_n}g(x)]
\eea
Thus the recipe for formulating the noncommutative version of the field theories is to replace all the ordinary products by the
Moyal products. As can be seen, such theories are highly non-local and bring with them several new features like UV/IR mixing
\cite{shiraz} and unitarity problems \cite{gomis}. However, inspite of these there has been considerable activity in this field
\cite{activity}. The noncommutative version of quantum electrodynamics has been examined in \cite{ncqed}.
A method for formulating the non-abelian noncommutative field theories has been discussed in \cite{nonabelian}
and using these ideas noncommutative version of the standard model has been proposed \cite{ncsm}. In the noncommutative
version, there are several new features and interactions, like triple gauge boson vertices,
 that appear and some of the related phenomenological aspects of these have been investigated \cite{ncsmpheno}.
\end{section}
%%%%%%%%%%%%%%%%%%%%%%%%%%%%%%%%%%%%%%%%%%%%%%%%%%%%%%%%%%%%%%%%%%%%%%
\begin{section}{Noncommutative corrections to $t \to b W$}
\indent We take as our tarting point the action given in \cite{ncsm} for the quark sector. We asume here that the fields have
been redefined in terms of the physical fields and as far as this work is concerned, we only concentrate on the charged current
interactions. The action with the ordinary products replaced by Moyal products looks like
\be
S_{charged-current} = \frac{g}{2\sqrt{2}}\int d^4x \sum_{families}\Bigg(\bar{\nu}\ast\not{W}\ast(1-\gamma_5)e ~+~ 
V_{ud}\bar{u}\ast\not{W}\ast(1-\gamma_5)d\Bigg)
\ee
where we generically denote the leptons as $e$ and their corresponding neutrinos as $\nu$ and for the quarks, we use
$u$ and $d$ to denote the up and the down type quarks. $W$ is the W-boson field and $V_{ud}$ are the relevant CKM elements.
The fields are expanded in powers of $\Theta$ and here we do not concern ourselves here about 
the terms that have more than one gauge field coupling to
fermions. The fermion fields when expanded to ${\mathcal{O}}(\Theta^2)$ have the following structure
\be
\psi = \Bigg(1 - \frac{1}{2}g\Theta^{\mu\nu}A_{\mu}\partial_{\nu} - \frac{i}{8}g\Theta^{\mu\nu}\Theta^{\rho\sigma}\partial_{\mu}
A_{\rho}\partial_{\nu}\partial_{\sigma}\Bigg)\psi_0 + {\mathcal{O}}(\Theta^3)
\ee
where $\psi_0$ refers to the usual fermionic field, $A$ is the associated gauge field and $g$ is the gauge coupling.
Plugging this expression in the action, we get the matrix element for the process $t(p_t) \to b(p_b) + W(p_W)$
\bea
{\mathcal{M}} &=& \Bigg(\frac{ig}{2\sqrt{2}}V_{tb}\Bigg)\bar{b}(p_b)\Bigg[\gamma^{\beta} - \frac{1}{2}(\Theta^{\mu\beta}
\not{p_t}{p_W}_{\mu} + \Theta^{\beta\alpha}\not{p_W}{p_t}_{\alpha} - 
\Theta^{\mu\alpha}\gamma^{\beta}{p_W}_{\mu}{p_t}_{\alpha}) \\ \nonumber
&+& \frac{i}{8}(\Theta^{\mu\beta}\gamma^{\alpha} + \Theta^{\beta\alpha}\gamma^{\mu} - \Theta^{\mu\alpha}\gamma^{\beta})
\Theta_{\rho\sigma}{p_W}_{\alpha}{p_t}_{\mu}{p_W}^{\rho}{p_t}^{\sigma}\Bigg](1-\gamma_5)t(p_t)\epsilon^{\ast}_{\beta}(p_W)
\eea
The charged current contribution due to noncommutativity in our case is same as that obtained by Iltan \cite{ncsmpheno} in
context of W decay into a lepton and anti-neutrino modulo an overall sign factor arising due to difference in the convention
for momentum flow. The noncommutative correction to the matrix element explicitly reads (after some algebra)
\bea
{\mathcal{M}}_{NC} &=& \Bigg(\frac{ig}{2\sqrt{2}}V_{tb}\Bigg)\epsilon^{\ast}_{\beta}\Bigg[\bar{b}(p_b)\gamma^{\beta}(1-\gamma_5)t(p_t)
\Bigg(\frac{1}{2}\Theta^{\mu\alpha}{p_W}_{\mu}{p_t}_{\alpha} + \frac{i}{8}\Theta^{\rho\sigma}{p_W}_{\rho}{p_t}_{\sigma}\Bigg)\\ \nonumber
&-& \frac{m_t}{2}\bar{b}(p_b)(1+\gamma_5)t(p_t)~\Theta^{\beta\lambda}{p_b}_{\lambda} + 
\frac{m_b}{2}\bar{b}(p_b)(1-\gamma_5)t(p_t)~\Theta^{\beta\delta}{p_t}_{\delta} \Bigg] \label{eq.1}
\eea
\indent Using the expression for the matrix element above, the decay rate can be easily evaluated. The decay rate, including
the noncommutative effects, can be expressed in the following form
\be
\Gamma_{Total} = \frac{\vert V_{tb}\vert^2}{16\pi p_t^0}\Bigg(\frac{g}{2\sqrt{2}m_W}\Bigg)^2
\lambda^{\frac{1}{2}}\Bigg(1,\frac{m_W^2}{m_t^2},\frac{m_b^2}{m_t^2}\Bigg) [SM + NC_1 + NC_2]
\ee
where
\be
SM = 2[m_W^2(m_t^2 + m_b^2 - m_W^2) + (m_t^2 - m_b^2 - m_W^2)(m_t^2 - m_b^2 + m_W^2)]
\ee 
\bea
NC_1 &=& \Theta^{\alpha}_{\beta}\Theta^{\beta\sigma}{p_t}_{\alpha}{p_t}_{\sigma}\Bigg(\frac{1}{12m_t^2}\Bigg)\Bigg[2m_b^8 + 2m_t^8
- 3m_t^6m_W^2 + 3m_t^4m_W^4 + m_t^2m_W^6 \\ \nonumber
&-& 3m_W^8 - 5m_b^6(m_t^2 - m_W^2) + m_b^4(6m_t^4 - 7m_t^2m_W^2 - 19m_W^4) \\ \nonumber
&+& m_b^2(-5m_t^6 + 5m_t^4m_W^2 + 3m_t^2m_W^4 + 15m_W^6)\Bigg]
\eea
\be
NC_2 = -\Theta_{\alpha\beta}\Theta^{\beta\alpha}\Bigg(\frac{m_W^2}{24}\Bigg)\Bigg[(5m_b^2 + m_t^2 - m_W^2)[m_b^4 +
 (m_t^2 - m_W^2)^2 - 2m_b^2(m_t^2 + m_W^2)]\Bigg]
\ee
It is worthwhile to point out that the term linear in $\Theta$ that would arise due to the interference between the SM
and the noncommutative contributions does not show up in the decay rate. This is not hard to see because the phase space integrals
would yield terms proportional to the metric tensor or the top momentum and such terms would identically vanish because of
the antisymmetry of $\Theta^{\mu\nu}$.\footnote{ This is typical of a particle decaying and is not expected in a scattering process where
 the linear term will in general be present and give the leading contribution.}
 Thus, the decay rate simply splits into pure SM and NC contributions. It is well known that QCD corrections lower the tree level
decay rate by roughly $10\%$. From the above expressions we see that the contribution $NC_2$ has a negative sign and it is interesting
to see the effects of noncommutativity on the decay rate. We consider three cases: (a) when $\Theta^{0i}~=~0$,
 (b) when $\Theta^{ij}~=~0$ and (c) both  $\Theta^{0i}$ and $\Theta^{ij}$ non-zero. Noncommutativity in the time direction is
expected to lead to non-unitary behaviour of the theory but it has been pointed out by Liao and Sibold \cite{activity} that
careful handling of time ordered products and the time derivatives arising due to Moyal/star product avoids any such
problem. We thus retain the piece with noncummutativity in the time direction as well.  We define the following:
\be
\Theta^{0i} = \theta_{ts}^i \hskip 1.5cm \theta_{ss}^i = \frac{1}{2}\epsilon^{ijk}\Theta^{jk}
\ee
Also for the sake of simplicity
we assume that $\vert\theta_{ts}^{i}\vert~=~\vert\theta_{ss}^{i}\vert$ and denote it by $\Lambda_{NC}^{-2}$, where $\Lambda_{NC}$ (in 
the units of GeV) is the characteristic scale of noncummutative interactions. \\ \\
\indent In Fig.1, we plot the decay rate as a function of $\Lambda_{NC}$ for all the three cases. Only for the case when
the noncommutativity in the time direction is put to zero, we get a lower rate when the noncommutative effects are included.
But this effect is seen only for low values of $\Lambda_{NC}$ and as it is increased the rates in all the three cases approach
the SM value. The ratio of the decay rate for a longitudinally polarized W (with polarization vector 
$\epsilon_L^{\mu}~\sim~p_W^{\mu}/m_W$) to the unpolarized decay rate has been measured by 
the CDF collaboration \cite{cdf} and it comes out to be
\be
F_0 = \frac{\Gamma(t \to b W_L)}{\Gamma^{Total}(t \to b W)} = 0.91 \pm 0.37 \pm 0.13 \label{eq.1}
\ee
while the SM value is $F_0^{SM}~=~0.701$. It is straight forward algebra to get the noncommutative correction to the polarized rate 
\be
\Gamma_{NC}^{long} =  \frac{\vert V_{tb}\vert^2}{16\pi p_t^0}\Bigg(\frac{g}{2\sqrt{2}m_W}\Bigg)^2
\lambda^{\frac{1}{2}}\Bigg(1,\frac{m_W^2}{m_t^2},\frac{m_b^2}{m_t^2}\Bigg)\Delta_{long}
\ee
where 
\bea
\Delta_{long} &=& \frac{1}{12}[m_b^4 + (m_t^2 - m_W^2)^2 - 2m_b^2(m_t^2 + m_W^2)] \\ \nonumber
&&[2m_b^4 + 2m_t^2(m_t^2 - m_W^2) - m_b^2(m_t^2 + 2m_W^2)] \vert\Theta^{0i}\vert^2
\eea
One finds that the polarized rate contains no terms proportional to $\Theta^{ij}$ i.e. the polarized rate
only depends on $\Theta^{0i}$. Therefore in teh absence of space-time noncommutativity, there are no noncommutative corrections to the 
longitudinal polarized rate and the sole contribution comes from the SM.
Thus the quantity $F_0$ stands a good chance to determine the kind of
noncommutativity that is present in the theory.
In Fig.2 we plot $F_0$ against $\Lambda_{NC}$ and find that for lower values of the
noncommutativity scale, there is a significant deviation from the SM result for the case of zero time like noncommutativity
or when both are present. But as before, with the increase in $\Lambda_{NC}$, all the curves approach the SM value. We can try
to get some lower bound on the noncommutative scale by imposing the condition Eq.(\ref{eq.1}). If that is done, we get a bound
$\Lambda_{NC}~\geq~{\mathcal{O}}(100~GeV)$ for space-space noncommutativity while no physically meaningful values are obtained
in any of the other two cases. Infact, one gets imaginary values for the noncommutativity scale. This may be an indication that
a more careful and thorough treatment is required at the level of the action itself and it is hoped that such a treatment
should bypass any such troubles.  \\ \\
\indent The transverse plus rate $\Gamma_+$ is also of considerable interest because simple helicity
arguments lead to the conclusion that at the tree level this rate vanishes for vanishing $m_b$. Therefore, a non-vanishing
$\Gamma_+$ at the Born amplitude level can arise from $m_b \neq 0$ effects or can arise because of higher order corrections
beyond the Born amplitude. Also, such an effect can be an artifact of departure from the V-A current structure.
From the matrix element Eq.(\ref{eq.1}) it is clear that there is a departure from the V-A structure of the weak current
and this fact is expected to show up in corrections to $\Gamma_+$. We evaluate the noncommutative corrections to $\Gamma_+$.
The total rate is
\be
\Gamma_+ = \Gamma_+^{SM} + \Gamma_+^{NC}
\ee
where the SM expression is well known and the second term is
\be
\Gamma_+^{NC} = \frac{\vert V_{tb}\vert^2}{16\pi p_t^0}\Bigg(\frac{g}{2\sqrt{2}}\Bigg)^2
\lambda^{\frac{1}{2}}\Bigg(1,\frac{m_W^2}{m_t^2},\frac{m_b^2}{m_t^2}\Bigg)\Delta_+
\ee
where the quantity $\Delta+$ is
\bea
\Delta_+ &=& \frac{m_t^4}{24}(m_t^2 + m_b^2 - m_W^2)\sum_{i=1}^3\Theta^{0i}\Theta_{i0} - 
\frac{m_t^4}{48}(m_t^2 + m_b^2 - m_W^2)\sum_{a=1}^2\Theta^{a\lambda}\Theta_{\lambda}^a \\ \nonumber
&+& \frac{m_b^2m_t^2}{4}(m_t^2 + m_b^2 - m_W^2)\sum_{a=1}^2\vert\Theta^{a0}\vert^2 + 
\frac{m_b^2m_t^4}{2}\sum_{a=1}^2\vert\Theta^{a0}\vert^2
\eea
From this expression it is clear that even in the limit of vanishing $m_b$, the Born level noncommutative corrections to the
SM transverse plus rate are non-zero. Quite clearly, this stands out as another test of possible existence of any
noncommutativity. Also note that this rate picks out terms having noncommutativity in selected space directions as well
(the terms with index $a$). This fact can thus be used to distinguish between any inhomogeniety in the noncommutativity
parameter. However for the present analysis, we have assumed equal and constant values along all the directions but in
principle a more general analysis can be carried out using these expressions. The CDF measurement for the quantity 
$\Gamma_+/{\Gamma}$ is \cite{cdf}
\[
\frac{\Gamma_+}{\Gamma} = 0.11 \pm 0.15
\]
For the sake of illustration we consider the case of only space-space noncommutativity. Imposing the experimental results,
we get a central value for the noncommutativity scale as $\sim 50~GeV$. It is considered safe to say that if the top
decay reveals a deviation fromthe SM predictions at $1\%$ level, then the origin of such deviation
can be attributed to some non-SM physics. For the noncommutativity scale ${\mathcal{O}}(100)$ GeV, the estimated
value for $\Gamma_+/{\Gamma}$ is approximately (slightly less than) $0.01$ which satisfies the $1\%$ criterion and
therefore should be probed in future measurements. However,  
for higher values of the scale, the noncommutative corrections get smaller and smaller.\\ \\  
\indent Consider the radiative decay of the top quark, namely, $t\to b W \gamma$. 
The branching ratio for this decay channel (for a photon of energy > 10 GeV) is $3.5\times 10^{-3}$. The noncommutative 
corrections to this process come in the form of modification of the SM vertices plus a completely new vertex where
the radiated photon is attached to the tbW vertex. For the sake of illustration we consider this piece of the total amplitude.
The relevant interaction Lagrangian can be written in the following form:
\bea
{\mathcal{L}}_{tbWA} &=& \frac{eg}{2\sqrt{2}}\bar{b}(p_b)\Bigg[(\frac{1}{2}\Theta^{\mu\nu}\gamma^{\alpha}
+ \Theta^{\nu\alpha}\gamma^{\mu})\Bigg(-\frac{3}{4}(\partial_{\mu}A_{\nu} - \partial_{\nu}A_{\mu})W_{\alpha} \\ \nonumber
&-& (A_{\mu}W_{\nu} - A_{\nu}W_{\mu})\partial_{\alpha} + \frac{1}{6}(\partial_{\mu}W_{\nu} - \partial_{\nu}W_{\mu})A_{\alpha}
\Bigg)\Bigg](1-\gamma_5)t(p_t)
\eea 
The contribution of this piece alone for the bounds obtained above fits the experimental values and as expected,
for higher scales the noncommutative contribution diminishes and the SM piece surfaces to be the sole contributor.\\ \\ 
\indent The obtained bound is a very low bound and should be directly verifiable at the collider experiments. This value
is of the same order ($141~GeV)$ as obtained for the noncommutative QED (NCQED) for the process $e^+e^- \to \gamma\gamma$ at LEP by the OPAL
collaboration \cite{opal}. There is no reason that the NCQED and NCSM bounds be very similar but one can expect them
to be roughly of the same order. However, it is known that the collider bounds that one gets in case of NCQED are much lower than
the bounds obtained from Lamb shift measurements (Chaichian etal. \cite{ncqed}). Also, the siderial variation effects on
atomic clocks give very strong limits \cite{mocioiu}. Thus, we expect that more data and bounds
from other sources should supplement such priliminary results and give a better picture. 
Some more hints can be obtained by looking at the charged lepton (arising from the W decay) energy and angular spectrum.
This sector contains information about the top polarization as well which implies more independent data to confirm
the theoretical results. However, in this case it becomes even more complicated because even the W-decay element will
pick up additional noncommutative corrections and the whole picture gets very messy. Also, if one wants to use
the results obtained directly in a compleet analysis i.e. consider top production at a collider and then its decay etc,
 then these results have to be suitably convoluted with the SM expressions for other sub-processes to have a rough estimate
of the noncommutative corrections. A complete treatment would require a detailed analysis of noncommutative corrections
for each sub-process and a careful convolution to extract the leading terms.
Therefore, presently it appears that there is a very narrow window to probe noncommutative effects
by directly looking at the decay of the top quark as the SM contribution blinds the extra effect completely for almost all
the region of the parameter space. 
\end{section}
\begin{section}{Conclusions}
\indent In conclusion, the noncommutative effects seem to be completely hidden under the shadow of the SM results for
most part of the parameter space and its only in a very small range, for very low values of the
noncommutativity scale that there are any significant deviations from the SM values, both for the unpolarized rate and
$F_0$ and transverse plus rate. The suppression due to the noncommutativity scale is 
rather large ($\Lambda^{-4}$) to be easily overcome in this
case. Thus it may seem that there is not much hope in determining possible noncommutative effects directly from the
top quark decay. However, there is still hope as angular asymmetries and corelations between the W decay products, say
a lepton and an anti-neutrino, may still be able to reveal even the feeble presence of any terms not dictated by SM.
Therefore, a detailed analysis of the top production process followed by cascade decay studies can possibly
probe such effects. Moreover, the tbW vertex now contains terms that essentially give rise to right handed weak currents and
it should be possible to detect the presence of such terms either in hadronic scattering reactions (possibly at LHC)
or indirectly through their contribution to rare meson decays where we expect even the terms linear in $\Theta$
to contribute and give the dominant effect. Also, from a theoretical point of view, more careful treatment
at the level of action is required so as to avoid possible problems like encountering physically meaningless bounds.
\end{section}

%%%%%%%%%%%%%%%%%%%%%%%%%%%%
%\pagebreak

%%%%%%%%%%%%%%%%%%%%%%%%%%
%\pagebreak
\vskip 3cm
\begin{figure}[ht]
\vspace*{-1cm}
\centerline{
\epsfxsize=8cm\epsfysize=8cm
                    \epsfbox{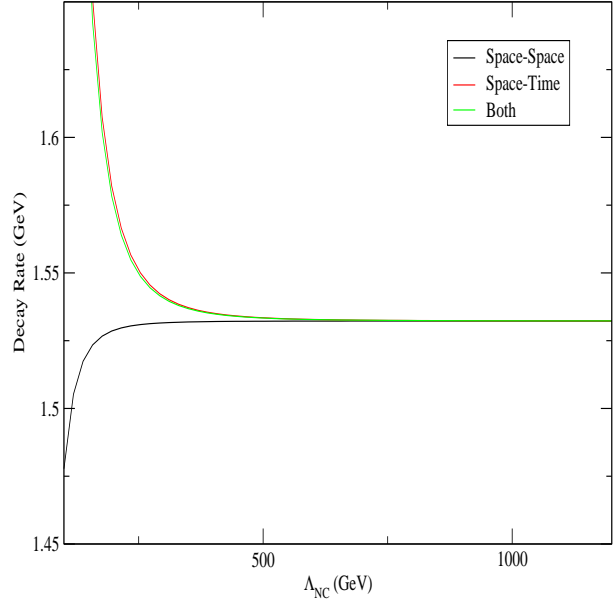}}
\caption{\em The decay rate (in GeV) as a function of the noncommutativity (in GeV) scale for space-space ($\Theta^{0i}=0$),
space-time ($\Theta^{ij}=0$) and both non-zero.}
\end{figure}
%%%%%%%%%%%%%%%%%%%%%%%%%%%%%%%%%5
\begin{figure}[ht]
\vspace*{-1cm}
\centerline{
\epsfxsize=8cm\epsfysize=8cm
                    \epsfbox{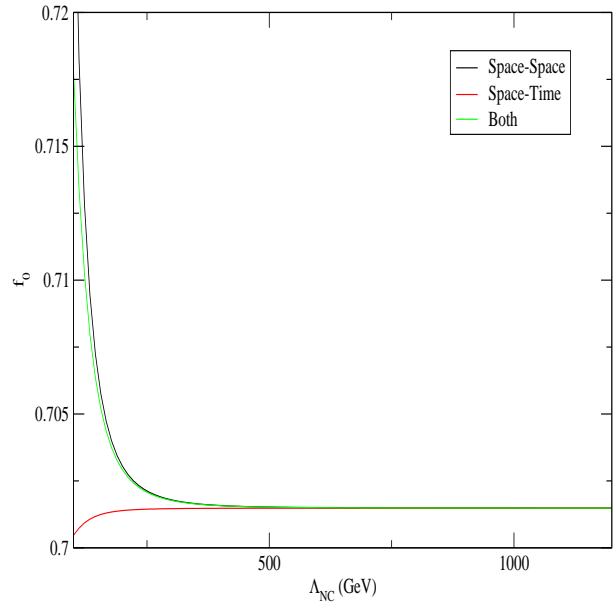}}
\caption{\em The quantity $F_0$ as a function of the noncommutativity (in GeV) scale for space-space ($\Theta^{0i}=0$),
space-time ($\Theta^{ij}=0$) and both non-zero.}
%	}
%label{fig:fig1}
\end{figure}
%%%%%%%%%%%%%%%%%%
\end{document}